\documentclass[review,12pt,3p]{elsarticle}

\usepackage[utf8]{inputenc}
\usepackage{amssymb}
\usepackage{graphicx}
\usepackage{amsmath}
\usepackage{amsfonts}
\usepackage{manfnt}

\usepackage{soul}
\usepackage{color}

\journal{Journal of Magnetism and Magnetic Materials}

\begin{document}

\title{Solutions to the Landau-Lifshitz-Gilbert equation in the frequency space: Discretization schemes for the dynamic-matrix approach}

\author[CBPF]{D. E. Gonzalez-Chavez}
\ead{diegogch@cbpf.br}
\author[CBPF]{G. P. Zamudio}
\author[CBPF]{R. L. Sommer}

\address[CBPF]{Centro Brasileiro de Pesquisas F\'{\i}sicas, 22290-180 Rio de Janeiro, RJ, Brazil}

\date{\today}

\date{\today}

\begin{abstract}
The dynamic matrix method addresses the Landau-Lifshitz-Gilbert (LLG) equation in the frequency domain by transforming it into an eigenproblem.
Subsequent numerical solutions are derived from the eigenvalues and eigenvectors of the dynamic matrix.
In this work we explore discretization methods needed to obtain a matrix representation of the dynamic operator, a fundamental counterpart of the dynamic matrix.
Our approach opens a new set of linear algebra tools for the dynamic matrix method and expose the approximations and limitations intrinsic to it.
Moreover, our discretization algorithms can be applied to various discretization schemes, extending beyond micromagnetism problems.
We present some application examples, including a technique to obtain the dynamic matrix directly from the magnetic free energy function of an ensemble of macrospins, and an algorithmic method to calculate numerical micromagnetic kernels, including plane wave kernels.
We also show how to exploit symmetries and reduce the numerical size of micromagnetic dynamic-matrix problems by a change of basis. 
This procedure significantly reduces the size of the dynamic matrix by several orders of magnitude while maintaining high numerical precision.
Additionally, we  calculate analytical approximations for the dispersion relations in magnonic crystals.
This work contributes to the understanding of the current magnetization dynamics methods, and could help the development and formulations of novel analytical and numerical methods for solving the LLG equation within the frequency domain.
\end{abstract}
\maketitle

\section{Introduction}

The Landau-Lifshitz-Gilbert (LLG) equation is the basis to the understanding of magnetization dynamics. This equation provides invaluable insights into the behavior of spins in response to external magnetic fields, paving the way for numerous technological advancements in the fields of spintronics, magnonics, and beyond. 
In modern spintronic \cite{RevModPhys.80.1517} and magnonic \cite{JourPhys:CondMatt.33.413001, IEETransMag.58.0800172} devices, magnetic materials oscillate in the gigahertz frequency range and sub-micrometer wavelengths.
These oscillations, known as spin waves are the basic foundation of several promising technologies in communication and computing devices, including magnonic crystals, spin-wave waveguides, spintronic oscillators, etc.
The LLG equation serves as a fundamental bridge between theory and experiment.
In particular, the frequency domain approach to the LLG equation allows for a detailed examination of the spin wave characteristics and their interaction with external fields and other material parameters.

Analytical solutions for the LLG equation, around an equilibrium position, in the frequency space have been obtained for several magnetic systems, including bulk magnetic materials \cite{PhysRev.73.155, PhysRev.110.1295, JMMM.248.164}, thin films \cite{J.Phys.Chem.Solids.19.308, JourAppPhys.41.987, SovPhysJour.24.718, JMMM.139.263, JMMM.163.39, PhysRevB.94.134408}, magnetic slabs \cite{JMMM.323.2418, PhysRevB.66.132402, PhysRevB.92.104424} and vortices \cite{AppPhysLett.81.1261, PhysRevLett.94.027205, PhysRevB.71.104415, PhysRevB.71.144407, PhysRevLett.122.097202}, among others.
A very important solution is the macrospin approximation, widely used for thin films and multilayered devices. 
This approximation is usually used to analyze or explain experimental data, including magnetic anisotropies \cite{JMMM.346.1, PhysRevB.88.104431}, damping \cite{JMMM.548.168923}, spin rectification \cite{JMMM.560.169614, JourApplPhys.123.123904}, magnetoimpedance \cite{JMMM.296.0304-8853}, and several other effects.
For an elaborate geometry, solutions in the frequency space must be obtained by numerical methods. 
These include the discretization of fields and operators involved in the LLG equation, and expressing the LLG equation in terms of a tensor formulation of the static and dynamic effective ﬁelds \cite{JMMM.248.164}.
The problem is finally formulated and numerically solved as an eigenvalue problem, using the method known as the dynamic-matrix approach \cite{PhysRevB.70.054409, JourCompPhys.228.6130, PhysRevB.70.184410}.
Over the last years, this method have been improved, and applied to several problems, including: Simulation of magnetic thermal noise \cite{JMMM.475.408}, spin wave propagation \cite{arXiv.1611.06153, NewJournPhys.21.033026, AIPAdv.11.095006, AIPAdv.12.115206}, analysis and separation of magnetic energy contributions \cite{PhysRevB.104.174414}, and other applications \cite{JMMM.546.168683, JourAppPhys.133.033902}.

In this work we explore discretization methods needed to obtain a matrix representation of the dynamic operator, which serves as a fundamental counterpart to the dynamic matrix.
Our approach opens a new set of linear algebra tools for the dynamic matrix method.
Using the fact that an (approximate) matrix representation of an operator can be obtained using any base of functions, we show an algorithmic way of calculating kernel matrices and the dynamic matrix.
Using this very same  method, we are able to obtain the dynamic matrix for an ensemble of macrospins directly from its free energy function.
Moreover, our approach clarifies the applicability of linear algebra tools to the dynamic-matrix problem.
This is demonstrated with examples of symmetry analysis and change of basis.
Using the method presented here, we achieved a significant reduction in the numerical problem's size by several orders of magnitude, while preserving high numerical precision. 
Additionally, we effectively exploit discrete symmetries, such as spatial parity, to separate the numerical problem even before solving it.
Furthermore we expose the approximations and limitations intrinsic to discretization in the dynamic-matrix method.

This work contributes to the understanding of the current magnetization dynamics methods, and could help the  development and formulations of novel analytical and numerical methods for solving the LLG equation within the frequency domain.
In particular, for numerical applications, methods that result in significant reductions in computational time are provided, along with the introduction of new tools for symmetry analysis.

This manuscript is organized as follows: 
In Sec. \ref{sec:II} we present a review of the overall theory of the dynamic magnetization in the frequency space, in terms of  integro-diferential operators. 
We also show how to obtain physical solutions for both free and forced oscillation problems around a magnetic equilibrium position, relying on the eigensolutions of the dynamic operator.
Then, in Sec. \ref{sec:III} we introduce a novel approach for discretization of the dynamic operator using any base of functions. 
We focus on the micromagnetic discretization, i.e. in terms of a grid or a mesh, and illustrate the well known procedure of reducing the numerical complexity of the system by rotating to the vector basis locally perpendicular to the equilibrium magnetization.
We move forwards and demonstrate how, via a general change (and reduction) of basis to any set of functions, the problem can be further simplified and separated by exploiting its symmetries.
Additionally, we show how the described methods can be used to derive analytical solutions for the dynamic-matrix problem using macrospin like approximations.
Furthermore, in Sec. \ref{sec:IV} we apply the presented methods to established problems, demonstrating improved methodologies for solution derivation and innovative analysis techniques.
In Sec. \ref{sec:macrospins}, we show how to derive the dynamic matrix for ensembles of macrospins directly form the free energy function expressed in terms of the magnetic moments that constitute the system. 
We include an example of results obtained by this method, and compare these to experimental measurements.
In Sec. \ref{sec:demag_kernels}, we use an algorithmic procedure to calculate micromagnetic kernels for a grid discretization and for mixture of plane waves and position-wise functions. 
Using the former kernel we find the dispersion relations and oscillation profiles of plane waves in a thin film.
In Sec. \ref{sec:fmr_problem}, we reproduce the proposed FMR problem for micromagnetic simulations and by employing a set of Legendre polynomials for a change of basis we reduce the size of the numerical problem and exploit the symmetries of the system.
Finally, in Sec. \ref{sec:magnonic_crystal} we use macrospin like approximations to obtain semi-analytical approximations for the dispersion relations in one dimensional magnonic crystals.
All the software implemented for these examples is available through Dymas \cite{dymas}, an open-source Python package for magnetization dynamics in the frequency domain.

\section{Magnetization Dynamics in the frequency space}
\label{sec:II}

The dynamics of the magnetization vector $\mathbf{M} = M_s \mathbf{m}$, where $M_s$ denotes the saturation magnetization and $\mathbf{m}$ is a unit vector, is described by the reduced Landau-Lifshitz-Gilbert (LLG) equation,
\begin{equation}
    \partial_t \mathbf{m} = 
    -\gamma \mathbf{m} \times \mathbf{H} 
    + \alpha \mathbf{m} \times \partial_t \mathbf{m},
\label{eq:LLG}
\end{equation}
where $\mathbf{H}$ is the effective field. 
It should be noticed that, typically, $-\mathbf{H} \cdot \mathbf{M}$ does not represent the magnetic energy density $e_m$.
Instead, the relation holds as $\mathbf{H} = -\tfrac{\delta e_m}{\delta \mathbf{M}}$.
Although not all magnetic self-interactions result in effective-field terms linear in $\mathbf{M}$, such terms can be linearized near an equilibrium position.
In general, $\mathbf{H}$ can be expressed as a Zeeman like field $\mathbf{H^\mathrm{Z}}$ plus terms that depends linearly on the $\mathbf{M}$ field. 
Given the linearity of $\mathbf{H}$ with $\mathbf{M}$, the Schwartz kernel theorem \cite{hormander1} ensures the existence of matrix function $\widehat{K}{(\mathbf{x},\mathbf{y})}$ such that $\mathbf{H}{(\mathbf{x},t)}$ at position $\mathbf{x}$ and time $t$ is given by Eq. \ref{eq:kernel},
\begin{equation}
    \mathbf{H}{(\mathbf{x},t)} 
    = \mathbf{H^\mathrm{Z}}{(\mathbf{x},t)} + 
    \int_V \widehat{K}{(\mathbf{x},\mathbf{y})} \, \mathbf{M}{(\mathbf{y},t)} d^3y
\label{eq:kernel}
\end{equation}
where the integral is performed over the position $y$ in the volume $V$ that encloses the magnetic system.
$\widehat{K}$ depends exclusively on the geometry of the system and the interactions of $\mathbf{M}$ with itself (demagnetization and exchange) or with the lattice (anisotropy).
$\widehat{K}$ can be calculated as a linear combination of matrix functions corresponding to the energy terms of the system. 
As such, $\mathbf{H}$  can be presented as the sum of field contributions from the interactions present in the system.

\subsection{Magnetization dynamics around an equilibrium position}

The time dependent $\mathbf{m}{(\mathbf{x},t)}$ field can be expressed as a $\mathbf{\delta m}{(\mathbf{x},t)}$ perturbation around an equilibrium field $\mathbf{m}^\mathrm{eq}{(\mathbf{x})}$
\begin{equation}
    \mathbf{m}{(\mathbf{x},t)} = 
        \mathbf{m}^\mathrm{eq}{(\mathbf{x})}
        + \mathbf{\delta m}{(\mathbf{x},t)}.
\label{eq:m_perturbation}
\end{equation}
where $\mathbf{m}^\mathrm{eq}$ and $\mathbf{\delta m}$ are perpendicular to each other, i.e. $\mathbf{\delta m}{(\mathbf{x},t)} \cdot \mathbf{m}^\mathrm{eq}{(\mathbf{x})} = 0$.
Furthermore, due to the equilibrium condition $\partial_t \mathbf{m}^\mathrm{eq} = 0$, the effective field at the equilibrium $\mathbf{H}^\mathrm{eq}{(\mathbf{x})}$ and $\mathbf{m}^\mathrm{eq}{(\mathbf{x})}$ are parallel to each other, locally, at all positions $\mathbf{x}$. 
With these conditions, the dynamics around the equilibrium position is described by  Eq. \ref{eq:Dynamics} and Eq. \ref{eq:dH}, where   $\mathbf{h}{(\mathbf{x}, t)}$ is the time dependent Zeeman field contribution that drives the magnetization out of equilibrium, and $\mathbf{\delta H}$ is the dynamic field produced by $\mathbf{\delta m}$ and $\mathbf{h}$.

\begin{equation}
   \partial_t \mathbf{\delta m}{(\mathbf{x},t)} 
 =  -\frac{\gamma}{1+\alpha^2{(\mathbf{x})}} 
    \Big[
       \mathbf{m}^\mathrm{eq}{(\mathbf{x})} 
        \times  
       \mathbf{\delta H}{(\mathbf{x}, t)} 
      - \alpha{(\mathbf{x})} \mathbf{m}^\mathrm{eq}{(\mathbf{x})} \times \big(
        \mathbf{\delta H}{(\mathbf{x}, t)}
        \times \mathbf{m}^\mathrm{eq}{(\mathbf{x})} \big)
    \Big]
\label{eq:Dynamics}
\end{equation}

\begin{equation}
  \mathbf{\delta H}{(\mathbf{x}, t)} 
  = \mathbf{h}{(\mathbf{x}, t)} + 
   \int_V 
    \left[ 
     M_{s}(\mathbf{y})\widehat{K}{(\mathbf{x},\mathbf{y})} 
     -  \mathbf{m}^\mathrm{eq}{(\mathbf{y})} \cdot \mathbf{H}^\mathrm{eq}{(\mathbf{y})} 
     \delta{(\mathbf{x}-\mathbf{y})}
    \right]
   \mathbf{\delta m}{(\mathbf{y},t)} d^3y \\
\label{eq:dH}
\end{equation}

From Eq. \ref{eq:Dynamics} is easy to see that, as expected,  $\partial_t \mathbf{\delta m}$ lay on the plane perpendicular to $\mathbf{m}^\mathrm{eq}$.
Furthermore, only the components of $\mathbf{\delta H}$ in this plane will be relevant to the magnetization dynamics.
These facts can be used to write this equation in terms of the operator $\widehat{S} = \mathbf{m}^\mathrm{eq} \times$ and the projection perpendicular to $\mathbf{m}^\mathrm{eq}$ operator $\widehat{P} = -\widehat{S}^2$ as:
\begin{equation}
   \widehat{P} \, \partial_t \mathbf{\delta m} 
 = \partial_t \mathbf{\delta m} 
 =  -\frac{\gamma}{1+\alpha^2} 
 (\widehat{I} + \alpha \widehat{S}) \widehat{S}  
 \, \widehat{P} \mathbf{\delta H}
 \label{eq:LLG_asOperators}
\end{equation}
where $\widehat{I}$ denotes the identity operator.
For convenience, we will also write:
\begin{equation}
\widehat{L} = -\frac{\gamma}{1+\alpha^2} 
 (\widehat{I} + \alpha \widehat{S}) \widehat{S}  
\end{equation}
and
\begin{equation}
\mathbf{\delta H} =  \mathbf{h} + \widehat{N} \mathbf{\delta m}
\end{equation}
where $\widehat{N} \mathbf{\delta m}$ is the rightmost term in Eq. \ref{eq:dH}. We also define the dynamic operator $\mathfrak{D}$ as:
\begin{equation}
    \mathfrak{D} = \widehat{L}\,\widehat{P}\,\widehat{N}
\label{eq:Decom_D}
\end{equation}

\subsubsection{Free oscillations}
\label{sec:FreeOsc}

For a static Zeeman field, i.e. $\mathbf{h}{(\mathbf{x}, t)} = 0$, the time derivative of $\mathbf{\delta m}$ can be written as the linear operator $\mathfrak{D}$ acting on $\mathbf{\delta m}$.
\begin{equation}
  \partial_t \mathbf{\delta m}{(\mathbf{x}, t)} =
    \mathfrak{D} 
    \mathbf{\delta m}{(\mathbf{x}, t)}
\label{eq:Linear_D}
\end{equation}
In this case, without any external excitation, a perturbation will decay back to the equilibrium position.
Given an initial condition $\mathbf{\delta m}(\mathbf{x}, t=0)$, 
solutions for $\mathbf{\delta m}{(\mathbf{x}, t)}$ are given in Eq. \ref{eq:m_t_solutions}, in terms of the eigenvalues $\lambda_r$ and eigenfucntions $\mathbf{f}_r$ of $\mathfrak{D}$ $(\mathfrak{D}\mathbf{f}_r = \lambda_r \mathbf{f}_r)$,
\begin{equation}
\begin{split}
    \mathbf{\delta m}{(\mathbf{x}, t)} & =
    e^{\mathfrak{D} t}  \mathbf{\delta m}{(\mathbf{x}, t=0)}  \\
    & = \sum_r{
    \langle \mathbf{\widetilde{f}}_r(\mathbf{x}), 
     \mathbf{\delta m}(\mathbf{x}, t=0)  
    \rangle
    e^{\lambda_r t} \mathbf{\delta m}_r{(\mathbf{x})}}
\label{eq:m_t_solutions}
\end{split}
\end{equation}
with $\langle a, b \rangle = \int_V (a^* \cdot  b) d^3 x$ denoting the inner product, and $\mathbf{\widetilde{f}}_r$ are the functions such that $\langle \mathbf{\widetilde{f}}_i , \mathbf{f}_j \rangle = \delta_{ij}$,  where $\delta_{ij}$ is the unit-less Kronecker delta.

Of course, this method works when we are able to solve the eigenvalue problem for $\mathfrak{D}$. Analytical  solutions for the eigenvalue problem of $\mathfrak{D}$ are only know for very simplified systems.
As stated in the introduction we will outline a numerical procedure to deal with this general eigenvalue problem.

\subsubsection{Forced oscillations}
\label{sec:ForcedOsc}

Magnetization dynamics experiments usually consist in obtaining the response of the magnetic system to some time dependent excitation.
In this case, we seek to obtain the differential susceptibility tensor $\mathfrak{X}$ of the system
\begin{equation}
    \partial_t \mathbf{m}{(\mathbf{x},t)} 
    = \mathfrak{X} \, \partial_t \mathbf{h}{(\mathbf{x},t)}.
\label{eq:Chi}
\end{equation}
If the output $\mathbf{\delta m}$ responds with the same frequency as the input $\mathbf{h}$, i.e. $\mathfrak{X}$ is linear in the frequency domain, then  Eq. \ref{eq:Chi} can be expressed in the frequency space as:
\begin{equation}
    \mathbf{\delta m}{(\mathbf{x},\omega)} 
    = \mathfrak{X}_\omega \, \mathbf{h}{(\mathbf{x},\omega)},
\label{eq:Chi_f}
\end{equation}
with $\mathbf{\delta m}$ been the forced response around the equilibrium position $\mathbf{m}^\mathrm{eq}$, due to the driving field $\mathbf{h}$.
Using Eq. \ref{eq:Chi_f} into Eq. \ref{eq:LLG_asOperators} we obtain $\mathfrak{X}_\omega$ as:
\begin{equation}
    \mathfrak{X}_\omega = 
    \left[ 
     i \omega \widehat{I} - \mathfrak{D}
    \right]^{-1} 
    \widehat{L} \widehat{P}
\label{eq:Chi_sol}
\end{equation}
Furthermore, if $\mathfrak{D}$ has no degenerate eigenvalues, guaranteeing the linear independence of its eigenfunctions, then the solution for $\mathbf{\delta m}$ can be expressed in terms of these eigenfunctions.
\begin{equation}
    \mathbf{\delta m}{(\mathbf{x}, \omega)} =
     \sum_{r}{
      \frac{
        \langle 
        \mathbf{\widetilde{f}}_r(\mathbf{x}), 
        \widehat{L} \widehat{P}
        \mathbf{h}(\mathbf{x}, \omega)  
        \rangle
      }
      {i\omega - \lambda_r} 
      \mathbf{f}_r(\mathbf{x})
      }
\label{eq:forced_solutions}
\end{equation}
This equation relates the amplitude and relative phase of a forced oscillation with its driving field. With this information is possible to reproduce experimental results such as power absorption in broadband FMR \cite{PhysRevB.88.104431}, FMR linewidth in non-saturated states \cite{JMMM.548.168923}, spin rectification voltages \cite{JMMM.560.169614}, among others.

\section{Discretization of the dynamic operator}
\label{sec:III}

Up to now we have established the connection between eigensolutions of $\mathfrak{D}$ and physical quantities as free or forced oscillations. 
Here, we outline how to obtain a matrix representation of $\mathfrak{D}$. 
This matrix form enables the numerical determination of eigenvalues and eigenvectors.

The matrix representation of a linear operator is not other than information about how the operator acts on a base of functions. 
In general, this matrix representation can be obtained by choosing a set of linearly independent functions $\{b_i\}$, such that exists a set $\{\widetilde{b}_j\}$ that satisfies $\langle \widetilde{b}_j, b_i \rangle = \delta_{ij}$.
With this basis, the elements $O_{ij}$ of the matrix representation of $\widehat{O}$ can be calculated as:
\begin{equation}
    O_{ij} = \langle \widetilde{b}_i, \widehat{O} \,b_j \rangle
    \label{eq:K_discretization}
\end{equation}

In our formalism, it is convenient to use a basis that separate the Euclidean basis $\{\hat{e}_1, \hat{e}_2, \hat{e}_3\}$ of the vector space from a set of discretization functions $\{p_i(\mathbf{x})\}$ for the position, with $p_i:\mathbb{R}^3 \rightarrow  \mathbb{R}$, requiring $\int_V p_i(\mathbf{x}) p_j(\mathbf{x})  d^3 x = \delta_{ij}$.
In this case, the basis functions can be grouped in sets of 3 functions $\{p_i(\mathbf{x}) \hat{e}_1, p_i(\mathbf{x}) \hat{e}_2, p_i(\mathbf{x}) \hat{e}_3\}$, and if the set $\{p_i(\mathbf{x})\}$ has $n$ elements, then any operator can be represented as a $3 \times n \times 3 \times n$ array.

For calculating the dynamic matrix, the first step is to find an approximate representation the $M_s \widehat{K}$ operator.
\begin{equation}
    J_{aibj} = \langle \hat{e}_a p_i, 
    M_s \widehat{K} \,
    \hat{e}_b  p_j \rangle
    \label{eq:K_as_tensor}
\end{equation}
We have purposefully include $M_s$ in this equation as it can change over the position. For a uniform magnetic material $M_s$ can be factored out of the inner product.
Following Eq. \ref{eq:dH}, $\widehat{N}$ can be represented as:
\begin{equation}
    N_{aibj} = J_{aibj} - \delta_{ab}\int_V p_i(\mathbf{x})
    \mathbf{m}^\mathrm{eq}(\mathbf{x}) \cdot \mathbf{H}^\mathrm{eq}(\mathbf{x})p_j(\mathbf{x}) d^3x  
    \label{eq:N_as_tensor}
\end{equation}
A similar procedure, must be applied to $\widehat{S}$, $\widehat{P}$ and $\widehat{L}$.
Finally, the discretization for the $\mathfrak{D}$ operator is also a  $3 \times n \times 3 \times n$ array,
\begin{equation}
    D_{aibj} = \sum_{ckdl}{L_{aick} P_{ckdl} N_{dlbj}} 
    \label{eq:D_as_matrix}
\end{equation}
from which eigenvalues $\lambda_r$ and eigenvectors $\mathbf{f}_r$ in the $\{p_i \hat{e}_1, p_i \hat{e}_2, p_i \hat{e}_3\}$ basis can be obtained.
\begin{equation}
    \lambda_r \mathbf{f}_r = D \mathbf{f}_r 
    \label{eq:eigen_D_discretized}
\end{equation}
Eq. \ref{eq:eigen_D_discretized} can be solved by mapping D to a $3n \times 3n$ matrix and using traditional numerical matrix solvers.
From this operation, $3n$ eigenvalues $\lambda_r$ will be obtained.
But, for an adequate basis, only $2n$ eigenvalues are expected to be non-zero, as $\mathbf{m}^\mathrm{eq}$ is an eigenfunction of $\mathfrak{D}$ with zero eigenvalue, and in the eigensolutions of the matrix representation this pair will appear $n$ times.

We must remark that the procedure described in this section consistently yields a numerical solution. 
This holds true regardless of the specific set of base functions chosen for discretization. 
The accuracy of the numerical results in characterizing a physical system relies on the capability of the selected basis to accurately represent the eigenfunctions of the $\mathfrak{D}$ operator.

\subsection{Rotation to a basis locally perpendicular to $\mathbf{m}^\mathrm{eq}$}
\label{sec:BaseRotation}

For an uniform $\mathbf{m}^\mathrm{eq}$ or for position-wise basis functions $\{p_i\}$, with $p_i$ associated to a $\mathbf{x}_i$ space point, the matrix representation of $\widehat{S}$ can be written as:
\begin{equation}
    S_{aibj} = \sum_{c}{\epsilon_{acb} (\mathbf{m}^\mathrm{eq}(\mathbf{x}_i)
    \cdot \hat{e}_c})\delta_{ij}
\end{equation}
And, it is greatly simplified if we transform from the $\{\hat{e}_1, \hat{e}_2, \hat{e}_3\}$ basis to a $\{\hat{o}_1(\mathbf{x}_i), \hat{o}_2(\mathbf{x}_i)\}$ orthonormal basis of the vector space that is locally perpendicular to $\mathbf{m}^\mathrm{eq}(\mathbf{x}_i)$.
In this new basis, $\widehat{S}$ can be regarded as a $90^\circ$ rotation and thus is represented by a $2 \times 2$ matrix
\begin{equation}
    S(x_i) = S = 
    \begin{bmatrix}
      0 & 1 \\
      -1 & 0 
    \end{bmatrix}
    \label{eq:S_2x2}
\end{equation}

The transformation between both basis is done with the help of a $3 \times 2$ rotation/projection matrix $R(\mathbf{x}_i)$ that can be calculated from the cross products of $\{\hat{e}_1, \hat{e}_2, \hat{e}_3\}$ with $\mathbf{m}^\mathrm{eq}(\mathbf{x}_i)$ \cite{PhysRevB.104.174414}. $R(\mathbf{x}_i)$ can also be calculated from the two eigenvectors with corresponding non-zero eigenvalues of the $P(\mathbf{x}_i)$ matrix of the $\widehat{P}$ operator. 
\begin{equation}
    P(\mathbf{x}_i)_{ab} = 1 - (\mathbf{m}^\mathrm{eq}(\mathbf{x}_i)\cdot \hat{e}_a) (\mathbf{m}^\mathrm{eq}(\mathbf{x}_i)\cdot \hat{e}_b)
\end{equation}
Then, the $\widehat{L}$ operator can be represented as a $2 \times n \times 2 \times n$ array.
\begin{equation}
    L_{uivj} = \sum_{u,i,v,j} {-\frac{\gamma}{1+\alpha(\mathbf{x}_i)^2} 
 (\delta_{uv} + \alpha(\mathbf{x}_i) S_{uv}) S_{uv} \delta_{ij}}
 \label{eq:L_discretization}
\end{equation}
Finally, the reduced representation of $\mathfrak{D}$ is also a $2 \times n \times 2 \times n$ array
\begin{equation}
    D_{uivj} = \sum_{a,b} {L_{uivj} 
    R(\mathbf{x}_i)_{ua} N_{aibj} R(\mathbf{x}_i)_{vb} }
    \label{eq:reduced_D}
\end{equation}
from which the eigenvalues $\lambda_r$ and eigenvectors $\mathbf{f}_r$ in the $\{\hat{o}_1(\mathbf{x}_i), \hat{o}_2(\mathbf{x}_i)\}$ basis can be calculated.
From this operation, $\mathbf{f}_r$ vectors with $2n$ components will be obtained. 
Then, $\mathbf{\widetilde{f}}_r$ vectors can obtained from the inverse of the eigenvector matrix.
Using $R(\mathbf{x}_i)$ the eigenvectors can be mapped back to the Euclidean space, and numerical solutions for forced or free oscillations can be obtained using Eq. \ref{eq:m_t_solutions} and Eq. \ref{eq:forced_solutions}

It must be noticed that the procedure described here is based on the premise that  $\mathbf{m}^\mathrm{eq}$ is an eigensolution of $\mathfrak{D}$. 
Separating the space in $\{\hat{o}_1(\mathbf{x}_i), \hat{o}_2(\mathbf{x}_i)\}$ and $\{\hat{\mathbf{m}}^\mathrm{eq}(\mathbf{x}_i)\}$ components will also separate the eigensolutions, and thus this procedure only obtain solutions with non-zero eigenvalues.

\subsection{Change of basis}
\label{sec:Change_of_basis}

The main difficulty in the discretization process is calculating a matrix representation of the kernel with components given by Eq. \ref{eq:K_as_tensor}. 
Fortunately this has already been addressed in micromagnetism by discretizing the system space using a grid or a mesh and employing $\{p_i\}$ functions such as Dirac deltas or box functions, as well as using plane waves for the reciprocal space (see Sec. \ref{sec:demag_kernels} for further details). 
Using these discretization schemes, it is always possible to obtain a good representation of a physical system provided that a large set of plane waves are employed or a sufficiently fine grid or mesh is used.
Unfortunately, this usually implies that a large number of discretization elements is used, as consequence, the arrays or matrices involved in the numerical solution become very large and cumbersome to work with.
Furthermore, usual micromagnetic mesh, grid or plane waves discretizations does not take into account the possible symmetries of the system. 

Here, we present a new method to address these issues. 
Our approach involves a transformation to a new basis with controlled symmetry properties in the position functions, and optionally reduced size in the number of elements.
We look for a new basis in the form of $\{q_k(\mathbf{x}) \hat{e}_1, q_k(\mathbf{x}) \hat{e}_2, q_k(\mathbf{x}) \hat{e}_3\}$.
The transformation from the $\{p_i\}$ basis to the $\{q_k\}$ basis is done the aid of matrices $Q$ and $\widetilde{Q}$, as described in Eq. \ref{eq:base_change}.
\begin{equation}
    N_{akbl} = \sum_{ij}{
     \widetilde{Q}_{il} N_{aibj} Q_{jk}
    }
    \label{eq:base_change}
\end{equation}
where $Q_{jk} = \langle q_k, p_j \rangle$, and $\widetilde{Q}$ is the Moore-Penrose inverse of $Q$.
If both $\{p_j\}$ and $\{q_k\}$ basis are orthonormal is easy to show that $\widetilde{Q}_{jk} = Q_{jk}^*$, i.e. $\widetilde{Q} = Q^{\dagger}$.

With this procedure, we can choose any set of $q_k(\mathbf{x})$ functions with the desired properties and symmetries.
For instance, if a smooth functions are used, e.x. polynomials, then a good numerical solution for smooth eigenfunctions can be obtained with a less number of polynomial coefficients than the equivalent in a grid or mesh discretization.
Additionally, choosing particular symmetries in the $\{q_k\}$ functions, a particular subset of eigensolutions can be obtained. This is demonstrated, using a numerical example, in Sec. \ref{sec:fmr_problem}.

\subsection{Macrospin like approximations}
\label{sec:Macrosping_basis}

Instead of solving the dynamic matrix equation using a large matrix representation and numerical algorithms, it is sometimes desirable to obtain analytical solutions for the natural oscillation frequencies.
To accomplish this, we need to reduce the size of the matrices to a manageable scale.
For any system, this can be achieved by considering a set of $\{q_k\}$ functions such that a function of the form $a q_k \hat{e}_x + b q_k \hat{e}_y + c q_k \hat{e}_y$ closely approximates an eigenfunction solution of the dynamic operator.
This implies that $q_k$ represents the spatial profile of an oscillation mode, wherein the mode maintains constant ellipticity and phase throughout all space.
The matrix representation of the dynamic operator in the $\{q_k\}$ basis will then be a block diagonal matrix, where each block can be solved independently.

If the system has a constant equilibrium magnetization oriented along the $x$ axis, then, using $q_k$ as basis function, the damping less version of the eigenproblem equation is:
\begin{equation}
    i \omega \mathbf{\delta m} = -\gamma 
    \begin{bmatrix}
      \widetilde{N}_{yz}^* & \widetilde{N}_{zz} \\
      -\widetilde{N}_{yy} & -\widetilde{N}_{yz}
    \end{bmatrix}
    \mathbf{\delta m},
\end{equation}
with solutions $\omega = \gamma \sqrt{\widetilde{N}_{yy} \widetilde{N}_{zz} - |\widetilde{N}_{yz}|^2}$, 
where $\widetilde{N}_{yy} = \langle q_k \hat{e}_y, \widehat{N} q_k \hat{e}_y \rangle$, $\widetilde{N}_{zz} = \langle q_k \hat{e}_z, \widehat{N} q_k \hat{e}_z \rangle$ and $\widetilde{N}_{yz} = \langle q_k \hat{e}_y, \widehat{N} q_k \hat{e}_z \rangle$.
For using this approximation, if all conditions are meet, we need to supply a $q_k$ function and calculate the $\widetilde{N}$ values. 
Of course if $q_k$ is the constant function, then the described procedure leads to the macrospin approximation, with solutions given by the Kittel formulas.

\section{Applications}
\label{sec:IV}

\subsection{Macrospins system}
\label{sec:macrospins}

A system comprising one or more interacting macrospins is of great interest, particularly for its application in the analysis of experimental results obtained from thin-film-based devices.
In this type of system, the magnetic free energy function $E_\mathrm{free}(\mu_1, ..., \mu_n)$ is typically known in terms of the magnetic moments $\mu_i$ that constitute the system. 
Here, we demonstrate how to derive a matrix representation of the $\widehat{N}$ operator directly form the energy function.

We begin by selecting a discretization basis, denoted as  $\{ p_i \hat{e}_a \}$, such that 
\begin{equation}
    \mu_i 
    = |\mu_i| \mathbf{m}_i = |\mu_i|\sum_{a}{m_{ia} p_i \hat{e}_a}
\end{equation}
Here, $m_{ia}$ are the components of the direction vector for $\mu_i$ and $|\mu_i|$ is the magnitude of the magnetic moment $\mu_i$. 
The magnetic free energy of the system is given by
\begin{equation}
    E_\mathrm{free}(\mu) = - \dfrac{1}{2} \sum_{i,j} 
    \langle \mu_i, \widehat{K} \mu_j \rangle
    - \sum_{i} \langle \mu_i, \mathbf{H}^\mathrm{Z} \rangle
\end{equation}
where $\widehat{K}$ is the kernel operator, and $\mathbf{H}^\mathrm{Z}$ is the Zeeman like field.
The free energy can be expressed in terms of $m_{ia}$ as:
\begin{equation}
\begin{split}
    E_\mathrm{free} = 
    & -\frac{1}{2} \sum_{iajb} |\mu_i||\mu_j| m_{ia} m_{jb} 
         \langle p_i \hat{e}_a, \widehat{K}_{ij} p_j \hat{e}_b \rangle \\
    & - \sum_{ia} |\mu_i| m_{ia} \langle p_i \hat{e}_a, \mathbf{H}^\mathrm{Z} \rangle
\end{split}
\end{equation}
The inner product $\langle p_i \hat{e}_a, \widehat{K} \, p_j \hat{e}_b \rangle = K_{iajb}$ inside the first sum yields the matrix representation of $\widehat{K}$. Its components can be acquired by exploiting the symmetry of $\widehat{K}$, and computing second partial derivatives with respect to the $m$ coefficients, resulting in:
\begin{equation}
K_{iajb} =
    -\frac{1}{|\mu_i| |\mu_j|}
    \frac{\partial^2  E_\mathrm{free}}{\partial m_{ia} \partial m_{jb}}
\end{equation}
On the other hand, the components of the effective field $\mathbf{H}$ can be obtained from the first derivatives of  $E_\mathrm{free}$
\begin{equation}
    \langle p_i \hat{e}_a, \mathbf{H} \rangle
     = 
     -\frac{1}{|\mu_i|}
     \frac{\partial E_\mathrm{free}}{\partial m_{ia}}
\end{equation}
Furthermore, the product $\mathbf{m}^\mathrm{eq}_i \cdot \mathbf{H}^\mathrm{eq}_i$ can be calculated as:
\begin{equation}
    \mathbf{m}^\mathrm{eq}_i \cdot \mathbf{H}^\mathrm{eq}_i = 
    -\frac{1}{|\mu_i|} \sum_{a} m_{ia} \frac{\partial E_\mathrm{free}}{\partial m_{ia}}
\end{equation}
Finally, following Eq. \ref{eq:K_as_tensor} and Eq. \ref{eq:N_as_tensor} we get the matrix representation for the $\widehat{N}$ operator:
\begin{equation}
    N_{iajb} =  
    - \frac{1}{|\mu_i|} \frac{\partial^2 E_\mathrm{free}}{\partial m_{ia} \partial m_{jb}} 
      +\sum_{c}{\frac{m_{ic}}{|\mu_i|}  \frac{\partial E_\mathrm{free}}{\partial m_{ic}}
     \delta_{ij} \delta_{ab}}
     \label{eq:N_macrospin}
\end{equation}
where the derivatives must be evaluated at the equilibrium position  $\mathbf{m}^\mathrm{eq}$. 
Of course, for calculation of solutions for the magnetization dynamics, the dynamic matrix must be calculated following the procedures described in the previous section. 

\begin{figure}[htbp]
 \includegraphics{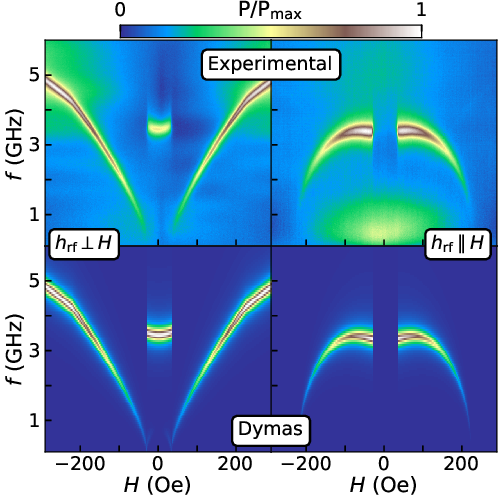}
 \caption{Experimental (top panels) and calculated (bottom panels) broadband FMR spectra for a SAF system for two different excitation field directions: $h_\mathrm{rf} \perp H$ (left panels) and $h_\mathrm{rf} \parallel H$ (right panels). Experimental details are available at \cite{JMMM.548.168923}, while calculation were performed using Dymas \cite{dymas} which implements the algorithms presented in this work.}
 \label{fig:macrospin}
\end{figure}

As a numerical example, we calculate the broadband FMR spectra, using Eq. \ref{eq:Chi_sol}, for a synthetic antiferromagnet (SAF) system and compare the results with experimental values.
The results are shown in Fig. \ref{fig:macrospin} and reveal a close agreement between the calculated and experimental results.
Details about the studied sample and the experimental setup for broadband FMR are provided in \cite{JMMM.548.168923}, while the energy description of the system is outlined in \cite{JMMM.560.169614}.
Our method and the Smith-Beljers approach applied to this system \cite{PhysRevB.88.104431,JMMM.560.169614} yields numerically equivalent dynamic matrices. 
In our approach, the dynamic matrix can be easily computed using Eq. \ref{eq:N_macrospin}, given that the magnetic free energy formula is known.
Notably, our method has the advantage over the Smith-Beljers approach as it does not involve singular points, making it easier to implement in software routines.

\subsection{Calculation of micromagnetic kernels}
\label{sec:demag_kernels}

Here, we demonstrate that the method presented in this work can be  algorithmically applied to obtain numerical micromagnetic kernels.
In particular, we show calculations for the conventional micromagnetic demagnetization kernel obtained through grid-like discretization, as well as the demagnetizing and exchange kernels associated with propagating spin waves in a film.

\subsubsection{Demagnetizing kernels for grids}

For the grid like discretization we use the standard basis for the Cartesian coordinate system $\{\hat{e}_a\}$ with $a \in \{1,2,3\}$ and box functions \mancube$_{i}$.
\begin{equation}
    \mbox{\mancube}_{i}(\mathbf{x}) = \frac{1}{\sqrt{V}}
      \Pi(\frac{x_{i1} - x_1}{\Delta_1})
      \Pi(\frac{x_{i2} - x_2}{\Delta_2})
      \Pi(\frac{x_{i3} - x_3}{\Delta_3})
\end{equation}
where $\Pi$ is the rectangular function, and $\mathbf{x}_i = (x_{i1}, x_{i2}, x_{i3})$, is the position vector of the center of a grid cell $i$, with volume $V = \Delta_1 \Delta_2 \Delta_3$. 
The discretization basis $\{\mbox{\mancube}_{i}\hat{e}_a\}$ has as many elements as 3 times the number of grid cells used for discretization.
This basis is orthonormal ($\langle \mbox{\mancube}_{i} \hat{e}_a , \mbox{\mancube}_{j} \hat{e}_b \rangle = \delta_{ij}\delta_{ab}$).
The matrix components $K_{iajb}^\mathrm{demag}$ of the demagnetizing kernel are:
\begin{equation}
    K_{iajb}^\mathrm{demag} = \langle \mbox{\mancube}_{i} \hat{e}_a , 
    \int \frac{-1}{4\pi}
     \nabla_{\mathbf{x}_i} \nabla_{\mathbf{x}_j} \frac{1}{|\mathbf{x}_i - \mathbf{x}_j|} \, 
    \mbox{\mancube}_{j} \hat{e}_b d^3x_j \rangle
    \label{eq:N_demag_grid}
\end{equation}
The term inside the integrals involved in this expression will be non-zero only inside the volumes $V_i$ and $V_j$ corresponding to the $i$ and $j$ cells . This leads us the expression:
\begin{equation}
    K_{iajb}^\mathrm{demag} = \frac{-1}{4\pi V} 
     \int_{V_i} d^3x_j 
     \int_{V_j} 
     \frac{\partial^2}{\partial x_{ia} \partial x_{jb}}
     \frac{1}{|\mathbf{x}_i - \mathbf{x}_j|} d^3x_j  
\end{equation}
These integrals are the same obtained by Newell et. al. \cite{JGeophysRes.98.9551}, which also calculated analytical solutions for them.

\subsubsection{Kernels for plane waves}
\label{sec:plane_waves_kernel}

Using the same method, we can calculate demagnetizing and exchange kernels for a film by combining plane waves in the plane directions of the film and $\Pi$ functions in the direction perpendicular to the plane.

We consider a film extended on the XY plane with side dimensions $L_x \times L_y$ and thickness $L_z$.
Here, it is convenient to label the basis elements $\{p_{i k_1 k_2}\}$ using the (discrete) index $i$ for denoting the discretization of the $[0,L_z]$ interval, and $k_1$, $k_2$ for the (continuous parameters) wave numbers in the plane of the film. This results in:
\begin{equation}
    p_{i \mathbf{k}}(x,y,z) = \frac{1}{\sqrt{L_x L_y} \sqrt{\Delta z}}\Pi(\frac{z-z_i}{\Delta z}) e^{i \mathbf{k}\cdot \boldsymbol{\rho} }
\end{equation}
where $\mathbf{k}= k_1 \hat{e}_1 +  k_2 \hat{e}_2$ and $\boldsymbol{\rho}=x \hat{e}_1 +  y \hat{e}_2$. 

For an infinite sample, i.e. in the limit where $L_x \to \infty$ and $L_y \to \infty$, the orthogonality of the complete basis reads as:
\begin{equation}
    \langle p_{i \mathbf{k}} \hat{e}_a, 
    p_{j \mathbf{k}'} \hat{e}_b \rangle = 
    \delta_{ab} \delta_{ij} \delta_{\mathbf{k},\mathbf{k}'} 
\end{equation}

To compute the matrix components of the demagnetizing kernel we need to follow the same procedure as in Eq. \ref{eq:N_demag_grid}, using the $p_{i \mathbf{k}}$ functions in this case.
Solution to the integrals involved in the inner product have been calculated by Guslienko et. al. \cite{JMMM.323.2418}.
Here, we expand their work to obtain the demagnetizing kernel including the discretization in the out of plane direction. The results for the demagnetizing kernel matrix components are summarized in Eq. \ref{eq:demag_k_1}, Eq. \ref{eq:demag_k_2} and Eq. \ref{eq:demag_k_3}, where $\alpha$ and $\beta \in \{1,2\}$ and $k=\sqrt{k_1^2+k_2^2}$.
Our results are consistent with those of Y. Henry et al. \cite{arXiv.1611.06153}. 
The advantage of the methods presented here, compared with previous approaches, lies in their straightforward applicability not only to demagnetizing kernels and plane waves but also to any operator representing a magnetic energy term and any set or combination of functions.

\begin{equation}
    K_{i\alpha j\beta \mathbf{k} \mathbf{k}'}^\mathrm{demag} = 
    \begin{cases}
        -\delta_{\mathbf{k},\mathbf{k}'} \frac{k_\alpha k_\beta}{k^2} e^{-k|z_i - z_j|} \Big[ \frac{\cosh{(k\Delta z)}-1}{k \Delta z} \Big] & ,  \text{if $i \neq j$} \\
        -\delta_{\mathbf{k},\mathbf{k}'} \frac{k_\alpha k_\beta}{k^2} \Big[ 1 + \frac{e^{-k\Delta z}-1}{k\Delta z} \Big]  & , \text{if $i=j$}
    \end{cases}
    \label{eq:demag_k_1}
\end{equation}
\begin{equation}
    -K_{i\alpha j3 \mathbf{k} \mathbf{k}'}^\mathrm{demag} = 
    K_{i3j\alpha \mathbf{k} \mathbf{k}'}^\mathrm{demag} = 
        -i \, \delta_{\mathbf{k},\mathbf{k}'} \frac{k_\alpha}{k} \, \mathrm{sign}(z_i - z_j) e^{-k|z_i - z_j|} \Big[ \frac{\cosh{(k\Delta z)}-1}{k \Delta z} \Big] \delta_{ij}
    \label{eq:demag_k_2}
\end{equation}
\begin{equation}
    K_{i3j3 \mathbf{k} \mathbf{k}'}^\mathrm{demag} = 
    \begin{cases}
        \delta_{\mathbf{k},\mathbf{k}'} e^{-k|z_i - z_j|} \Big[ \frac{\cosh{(k\Delta z)}-1}{k \Delta z} \Big] & ,  \text{if $i \neq j$} \\
        -\delta_{\mathbf{k},\mathbf{k}'} \Big[\frac{1-e^{-k\Delta z}}{k\Delta z} \Big]  & , \text{if $i=j$}
    \end{cases}
    \label{eq:demag_k_3}
\end{equation}

\begin{equation}
    K_{iajb \mathbf{k} \mathbf{k}'}^\mathrm{exch} = 
    \frac{2 A_\mathrm{ex}}{M_s}
    \left(
    \frac{\delta_{i,j-1} -2\delta_{ij} + \delta_{i,j+1}}{(\Delta z)^2}
     - k^2 \delta_{ij}
    \right)
     \delta_{ab} 
     \delta_{\mathbf{k},\mathbf{k}'}
    \label{eq:exch_k}
\end{equation}
    
The same procedure is applicable to the calculation of the exchange kernel matrix $K^\mathrm{exch}$. 
\begin{equation}
   K_{iajb \mathbf{k} \mathbf{k}'}^\mathrm{exch}
   =
   \langle p_{i \mathbf{k}} \hat{e}_a, 
   \frac{2 A_\mathrm{ex}}{M_s} \nabla^2
   p_{j \mathbf{k}'} \hat{e}_b \rangle
\end{equation}
In this case, solutions (see Eq. \ref{eq:exch_k}) are straight forward. For the used basis, the discretization along the Z axis naturally results into the three-term approximation to the second derivative \cite{PhysicaB.343.177}. Boundary conditions can be controlled by changing the properties of the $p_{i \mathbf{k}}$ functions at the top an bottom planes of the film.

It is noteworthy that both $K_{iajb \mathbf{k} \mathbf{k}'}^\mathrm{exch}$ and $K_{iajb \mathbf{k} \mathbf{k}'}^\mathrm{demag}$ include the term $\delta_{\mathbf{k},\mathbf{k}'}$.
This implies that these kernels are linear with respect to the wave vector. i.e. a magnetic excitation with a certain wavevector profile will generate an effective field with the same wave wavevector profile.
This arises from the translational symmetry of the system within the film plane, making plane waves eigenfunctions \cite{johnson2007notes} of the exchange and demagnetizing operators. 
While this is strictly applicable only to an infinite sample, it serves as a valid approximation for $k \ll L_x$ and $k \ll L_y$.
For uniform magnetization,  $\mathfrak{D}$ will also exhibit translational symmetry, leading to separable solutions in $k$.

\begin{figure}[htbp]
 \includegraphics{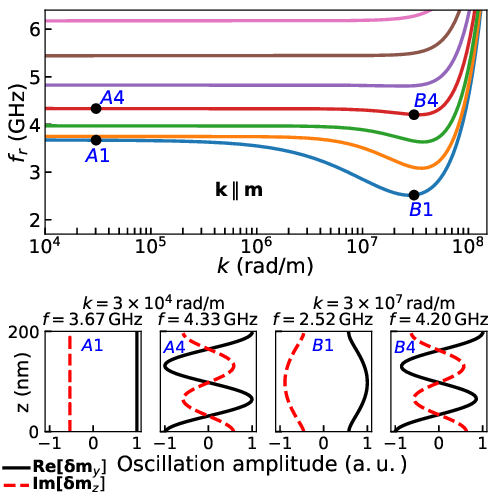}
 \caption{Top: Dispersion relations for spinwaves propagating in an in-plane magnetized film, with wave vector parallel to the magnetization. Each curve corresponds to an oscillation mode.
 Bottom: Oscillation profiles along the Z axis for the first and fourth modes, at the labeled points.}
 \label{fig:disp_rel_k}
\end{figure}

As a numeric example we obtain the dispersion relations for a 200 nm thick film, with $M_s = 140$ kA/m and $A_\mathrm{ex} = 3.6$ pJ/m, and an in-plane Zeeman field $\mu_0 H^Z = 70$ mT along the $\hat{e}_x$ axis.
Solutions for wavevectors  perpendicular ($\mathbf{k} = k \hat{e}_y$) to the magnetization are presented in Fig \ref{fig:disp_rel_k}. 
The Z axis was discretized in 50 elements, resulting in the obtainment of 100 eigenvalues for each $k$. 
For simplicity, only dispersion relations with positive frequency and below 6.3 GHz are presented.
The obtained dispersion relations, for the first oscillation modes, demonstrate a minimum at $k$ values around $3 \times 10^7$ rad/n this is a typical magnon frequency behavior for the $\mathbf{k} \parallel \mathbf{m}$ configuration, as utilized in Bose-Einstein magnon condensates experiments \cite{Nature.443.430, NewJourPhys.9.64}.
In Fig \ref{fig:disp_rel_k} we also present the mode profiles for the first and fourth modes at two different wavevectors $k =3 \times 10^4$ rad/m and $k= 3 \times 10^7$ rad/m. 
As expected, the precession around the equilibrium position consistently induces a phase difference of $\pi/2$ between $\mathbf{\delta m}_y$ and $\mathbf{\delta m}_z$. Specifically, in our findings, $\mathbf{\delta m}_y$ is purely real while $\mathbf{\delta m}_z$ is imaginary.
Moreover, the amplitudes along the $z$ position and the relative magnitudes of $\mathbf{\delta m}_y$ and $\mathbf{\delta m}_z$ vary across oscillation modes and along the dispersion relation.
These results allow for an analysis of the profile's dependence on $k$.
In particular, we observe that near the frequency minimum for the first mode, the ellipticity of the mode is close to 1 i. e. the amplitude of $\mathbf{\delta m}_y$ and $\mathbf{\delta m}_z$ are almost the same.
A detailed analysis of these numerical results will be published elsewhere.

\subsection{FMR standard micromagnetic problem using Legendre polynomials}
\label{sec:fmr_problem}

In this application example, we present solutions for the FMR  micromagnetic standard problem \cite{JMMM.421.428}. The studied system is  a permalloy cuboidal sample with dimension $120 \times 120 \times 10 \,\mathrm{nm}^3$, in equilibrium condition for a in plane Zeeman field with amplitude 80 kA/m and direction at 35$^\circ$ to the x-axis. 
Part of problem definition requires the analysis of the eigenmodes' resonance frequencies and spatial profiles. 

Obtaining solutions using the usual eigenvalue method is a straightforward application of the procedures described in this work. 
Here, we also explore the spatial symmetries of the system and reduce the size of the numerical problem. 
For this, we obtain a reduction of the  dynamic matrix calculated for the usual grid basis, applying the procedure described in Sec. \ref{sec:Change_of_basis}, using as new basis a combination of Legendre polynomials $P_n(x)$.

\begin{figure*}[htbp]
 \includegraphics{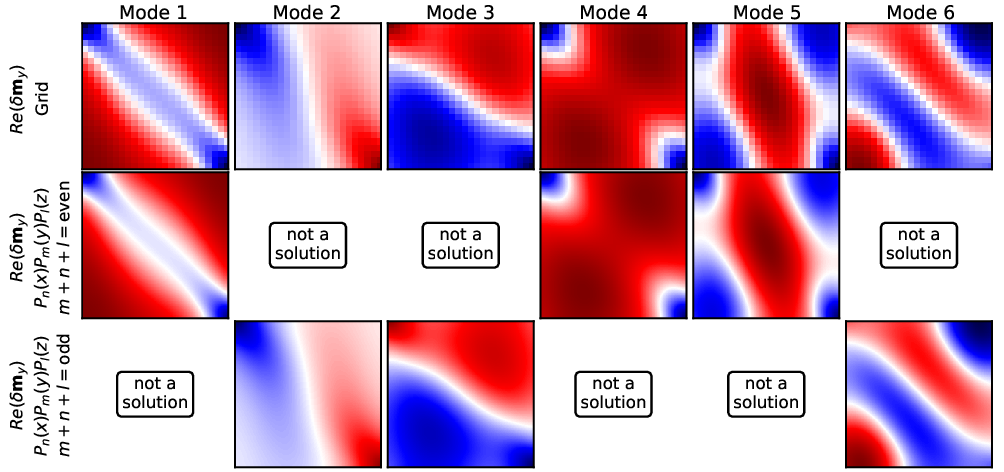}
 \caption{Spatial distribution of amplitude of the real part of the $y$ component of the oscillating magnetization ($\delta\mathbf{m}_y$) for different resonant modes, obtained for a grid discretization (top row), and using a combination of Legendre polynomials with even or odd symmetry (second and third row).}
 \label{fig:FMR_problem_modes}
\end{figure*}

For discretization in the \mancube$_i$ basis we use a $5 \times 5 \times 5 \, \mathrm{nm}^3$ cell size resulting into a $24 \times 24 \times 2$ grid. 
For the polynomial basis we use $q_k = P_{n_k}(x) P_{m_k}(y) P_{l_k}(z)$, with polynomial degrees $n$ and $m$ taken from 0 to 9 and $l$ been 0 or 1.
We also require that $q_k$ is either symmetric ($m+n+l=$ even) or anti-symmetric ($m+n+l=$ odd) with respect to the origin.
This results in two different basis with 100 elements each.

\begin{table}[htbp]
\begin{tabular}{c|c|c|c}
        & \multicolumn{3}{c}{Frequency (GHz)}                  \\
\hline
        & \hspace{1.5cm} & $P_n(x)P_m(y)P_l(z)$ & $P_n(x)P_m(y)P_l(z)$ \\
Mode \# & Grid   & n+m+l=even           & n+m+l=odd            \\
\hline
1       & 8.269  & -                    & 8.270                \\
2       & 9.408  & 9.408                & -                    \\
3       & 10.840 & 10.840               & -                    \\
4       & 11.237 & -                    & 11.238               \\
5       & 12.004 & -                    & 12.004               \\
6       & 13.057 & 13.057               & -                    \\
7       & 13.827 & -                    & 13.827               \\
8       & 14.289 & 14.289               & -                    \\
9       & 15.340 & -                    & 15.340               \\
10      & 15.934 & 15.934               & -                    \\
11      & 16.746 & 16.746               & -                    \\
12      & 17.258 & -                    & 17.258               \\
13      & 17.482 & -                    & 17.482               \\
14      & 18.442 & 18.443               & -                    \\
15      & 19.856 & -                    & 19.862              
\end{tabular}
\label{table:FMR_problem_freqs}
\caption{Calculated resonance frequencies of the system studied in the FMR micromagnetic standard problem, using a grid discretization and using a combination of Legendre polynomials with even or odd symmetry.}
\end{table}

We obtain 3 different sets of results:
The eigenvalue method for the grid (\mancube \, basis), and two for Legendre polynomials ($P_n(x) P_m(y) P_l(z)$ basis) with $m+n+l=$ even or odd. 
Results are summarized in Table \ref{table:FMR_problem_freqs} where the calculated resonant frequencies are presented, and in Fig \ref{fig:FMR_problem_modes} where the calculated resonant spatial profiles are shown.
The grid solutions, as expected, are very close to the values and profiles reported in the problem specification.
Solutions using the Legendre polynomials with $m+n+l=$ even are completely different from the solutions for the $m+n+l=$ odd set. 
Nevertheless, these are complementary and fully reproduce the grid solutions with high accuracy.
The separation of the solutions into two different classes is a consequence of the symmetry properties of the new basis. 
In this case, symmetries in the spatial profiles of the resonant modes, in each class, are the same of their corresponding basis. 
This can be evidenced by analyzing the sign of the plotted oscillation profiles, colored red for positive and blue for negative.
Solution for $m+n+l=$ even have the same sign in two opposite corners of the cuboid, while for $m+n+l=$ odd the signs are different.

The methods demonstrated here result in a significant decrease in dynamic matrix size, reducing computation time for determination of eigensolutions by orders of magnitude.
Specifically, in our example, the matrix shrinks from 5.3 million elements during grid discretization to just 40 thousand elements in the polynomial basis, all while maintaining result accuracy.
These efficiency enhancements are not limited to the use of Legendre polynomials but extend to other smooth functions capable of describing system symmetries.

It must be noticed that for applying the method described above, it is not strictly necessary to calculate the dense matrix $D$ in the grid base.
From Eq. \ref{eq:D_as_matrix} and Eq. \ref{eq:base_change} it is clear that the representation of the dynamic matrix in the new base is 
\begin{equation}
    D^q = \widetilde{Q} L P N Q,
    \label{eq:D_in_new_base}
\end{equation}
where $L$, $P$ and $N$ are calculated in the grid basis while $Q$ and $\widetilde{Q}$ define the transformations between basis.
We can separate Eq. \ref{eq:D_in_new_base} into two parts, the first one $\widetilde{Q} L P$ can be easily computed because $L P$ is a sparse matrix.
On the other hand, the complexity of calculating the second part $N Q$ is mainly due the needed calculation of $K Q$ which involves the micromagnetic kernel matrix. 
However, using the ideas described in \cite{JourAppPhys.133.033902}, we do no need to use the dense matrix $K$ to calculated $K Q$.
Instead, it can be obtained from the vectors resulting from $K q_k \hat{e}_a$, where $q_k$ are our new base functions and $\hat{e}_a$ are the unit vectors. 
And this operation of applying $K$ to arbitrary vectors ($q_k \hat{e}_a$ in this case) can be efficiently carried out by specialized algorithms as the traditional algorithms used in time domain micromagnetic software.
Other algorithmic optimizations are always possible, but are out of the scope of this work.

\subsection{One dimensional magnonic crystal}
\label{sec:magnonic_crystal}

In this final application example, we used the methods described in Sec. \ref{sec:Macrosping_basis} to obtain semi-analytical approximations for the dispersion relations in one dimensional magnonic crystals.
We consider a magnonic crystal magnetized along the $x$ axis and structured via a periodic modulation of the saturation magnetization $M_s$ along the $y$ axis.
In order to calculate the dynamic matrix of the system we need to obtain not only the kernel matrix $K$, but a matrix representation of $\widehat{N} = M_s \widehat{K}$.
To this end, it is convenient to use plane-wave functions.
Here, we only focus on waves propagating in the $y$ direction. 
Following the methods outlined in Sec. \ref{sec:demag_kernels}, we use orthonormal basis functions $p_k \propto e^{i k y}$. 
Using the Fourier series representation of $M_s$:
\begin{equation}
    M_s(y) = \sum_{n = -\infty}^{\infty} c_n e^{i\frac{2 \pi}{a}ny},
\end{equation}
where $a$ is the magnonic crystal lattice and  
\begin{equation}
    c_n = \frac{1}{a} \int_{-a/2}^{a/2} M_s(y) e^{-i\frac{2 \pi}{a}ny} dy,
\end{equation}
it is easy to see that $p_k M_s$ can be represented as a sum of $p_k$ functions.
Therefore, the matrix elements $N_{kalb} = \langle p_k \hat{e}_a, M_s \widehat{K} p_l \hat{e}_b \rangle $ are:
\begin{equation}
    N_{kalb} = \sum_{n = -\infty}^{\infty} c_n \langle p_{k-\frac{2 \pi}{a}n} \hat{e}_a, \widehat{K} p_{l} \hat{e}_b \rangle,
\end{equation}
were $\langle p_k \hat{e}_a, \widehat{K} p_l \hat{e}_b \rangle$ are the elements of the kernel matrix obtained using the plane wave functions.
The simple process described above is analogous to well-known methods for obtaining demagnetization kernel matrices involving a convolution of the kernel in the $k$-space with the Fourier transform of $M_s$.
It must be noticed that even for diagonal kernel matrices $N$ will not be a diagonal matrix. 
This implies, as expected, that the $p_k$ functions are not eigenfunctions of $N$.

In this example we go beyond finding the $N$ matrix for numerical solutions, instead we obtain analytical approximations for the dispersion relations of the oscillation modes.
We expect the profile of the oscillation modes to be described by a Bloch wave $q_k = f(y) p_k$ where $f(y)$ is a periodic function with Fourier series coefficients $b_l$.
Requiring $\langle q_k, q_k \rangle = 1$, and following the procedure described in Sec. \ref{sec:Macrosping_basis}, we obtain
\begin{equation}
    \widetilde{N}_{ab}(k) = 
      \sum_{n,m,l = -\infty}^{\infty}
       c_n b_{m}^* b_{l}
      \langle p_{k+\frac{2 \pi}{a}(m-n)} \hat{e}_a, \widehat{K} p_{k+\frac{2 \pi}{a}l} \hat{e}_b \rangle .
\end{equation}
Then, using the Parseval's theorem and the fact that the kernel matrix is diagonal in the plane wave representation, we obtain:
\begin{equation}
    \widetilde{N}_{ab}(k) = 
      \sum_{n = -\infty}^{\infty}
       b_n^* d_n
      \langle p_{k+\frac{2 \pi}{a}n} \hat{e}_a, \widehat{K} p_{k+\frac{2 \pi}{a}n} \hat{e}_b \rangle .
  \label{eq:N_effective_crystal}
\end{equation}
were
\begin{equation}
    b_n = \frac{1}{a} \int_{-a/2}^{a/2} f(y) e^{-i\frac{2 \pi}{a}ny} dy,
\end{equation}
and
\begin{equation}
    d_n = \frac{1}{a} \int_{-a/2}^{a/2} M_s(y) f(y) e^{-i\frac{2 \pi}{a}ny} dy,
\end{equation}
As explained in Sec. \ref{sec:Macrosping_basis}, from the values of $\widetilde{N}^{k}_{ab}$ we can calculate the natural frequency of the oscillation mode described by $q_k$.

We apply this approximation to a magnonic crystal composed of long rectangular stripes with width $w$ and height $h$, aligned along the $x$ axis and arranged parallel to each other in a plane, exhibiting a periodicity $a$ in the $y$ axis.
We consider only the demagnetizing kernel, no damping, magnetization equilibrium along the $x$ axis, and no external field.

\begin{figure}[htbp]
 \includegraphics{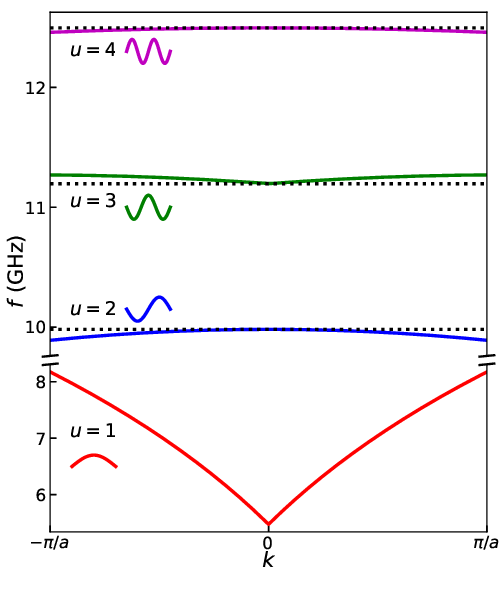}
 \caption{Calculated dispersion relations for the first four modes of a 1D magnonic crystal with lattice constant $a$, comprising parallel long slabs with width $0.9a$ and height $0.05a$. Corresponding oscillation profiles inside one of the slabs for each mode are labeled adjacent to the dispersion relations. Horizontal dotted lines serve as guides to identify the curvature of the dispersion relations.}
 \label{fig:magnonic_crystal}
\end{figure}

As $f(y)$ we use the periodically repeated eigen-solutions for a single stripe. 
We consider these to be harmonic functions with roots at the $y$ edges of the stripes.
These selections results in:
\begin{equation}
    b^{u=\mathrm{odd}}_n = \frac{\sqrt{2aw}}{\pi}
    \left( 
        \frac{\sin(\frac{au-2nw}{2a} \pi)}{au-2nw}
        +\frac{\sin(\frac{au+2nw}{2a} \pi)}{au+2nw}
    \right)
\end{equation}
for $u$ odd and $f(y) \propto \cos(\frac{\pi}{w}uy)$, 
\begin{equation}
    b^{u=\mathrm{even}}_n = \frac{\sqrt{2aw}}{i\pi}
    \left( 
        \frac{\sin(\frac{au-2nw}{2a} \pi)}{au-2nw}
        -\frac{\sin(\frac{au+2nw}{2a} \pi)}{au+2nw}
    \right)
\end{equation}
for $u$ even and $f(y) \propto \sin(\frac{\pi}{w}uy)$, and $d_n^u = M_0 b_n^u$, where $M_0$ is the saturation magnetization inside the stripes.
Furthermore, only $\widetilde{N}_{yy}$ and $\widetilde{N}_{zz}$ will be non-zero, which for the $u$th mode and $k$ wave number are:
\begin{equation}
    \widetilde{N}^{u}_{yy}(k) = 
      \sum_{n = -\infty}^{\infty}
       M_0 |b_n^u|^2 
       \left(
        1- \frac{1-e^{-(|k|+\frac{2 \pi}{a}n)h}}{(|k|+\frac{2 \pi}{a}n)h}
       \right)
    \label{eq:Nyy_crystal}
\end{equation}
and 
\begin{equation}
    \widetilde{N}^{u}_{zz}(k) = 
      \sum_{n = -\infty}^{\infty}
       M_0 |b_n^u|^2
      \left( 
        \frac{1-e^{-(|k|+\frac{2 \pi}{a}n)h}}{(|k|+\frac{2 \pi}{a}n)h}
      \right).
    \label{eq:Nzz_crystal}      
\end{equation}
The frequency solution to the $u$th mode is:
\begin{equation}
 2 \pi f^u(k) = \gamma \sqrt{\widetilde{N}^{u}_{zz}(k) \widetilde{N}^{u}_{yy}(k)}
\end{equation}
We found that the series presented in Eq. \ref{eq:Nyy_crystal} and Eq. \ref{eq:Nzz_crystal} converge very fast.
In order to obtain numerical values for the frequency, we have used the following values for the structural parameters: $M_0 = 800$ kA/m, $w = 0.9 a$ and $h= 0.05 a$. 
Results for the first four modes are presented in Fig \ref{fig:magnonic_crystal} for $k$ inside the first Brillouin zone.
We observed that the slope of the dispersion relations increases with $\lvert k \rvert$ for the odd modes (cosine profiles) and decreases with $\lvert k\rvert$ for the even modes (sine profiles). However, for higher-order modes, the dependence of $f$ on $k$ is reduced. 
The overall dispersion relation resembles the band structure reported in the literature for similar 1D magnonic crystals \cite{AppPhysLett.90.092503, AppPhysLett.92.132504, AppPhysLett.94.083112}.

\section{Final remarks and conclusions}

We have shown how the general discretization approach of the dynamic matrix applies to different problems.
In micromagnetic applications, the most promising procedures presented here involve the calculation of the dynamic matrix and kernel matrices using any set of basis functions. These methods offer significant reductions in the size of the numerical problem by several orders of magnitude and effectively exploit symmetries, including discrete symmetries.
The procedures outlined in Section \ref{sec:demag_kernels} introduce a novel method for formulating and numerically computing kernel matrices. These techniques extend beyond the showcased example of plane wave analysis, as they can be applied to any basis. 
This versatility allows for the exploitation of symmetries or the reduction of the size of the final matrix in various applications.
However, it is important to note that calculating the $D$ matrix requires not only the kernel matrix but also the matrix representation of $\widehat{S}$, which is determined by the magnetic equilibrium configuration.
Calculating this representation can be challenging for arbitrary basis functions unless the conditions described in Sec. \ref{sec:BaseRotation} are met.
To circumvent this issue, one can compute the $D$ matrix in the new basis, as demonstrated in Sec. \ref{sec:fmr_problem}.
Additionally, utilizing discrete symmetries not only reduces system complexity but also provides new analytical tools.

We have also presented in Sec. \ref{sec:macrospins} new algorithmic methods to solve the dynamics of ensembles of macrospins system starting from the free energy function in term of the constituting magnetic moments of the ensemble. 
This algorithm offers an advantage over traditional Smith-Beljers methods by avoiding encountering singular points.
As a result, simpler and more versatile algorithmic implementations can be employed.

From dynamic matrix theory, it is evident that the dynamic operator and the kernel operators do not necessarily share eigenfunctions.
However, under certain circumstances or approximations, the eigensolutions of the kernel can be employed for the calculation and analysis of oscillation modes.
This analysis can be extended to isolate the various energy contributions within the kernel and study their corresponding oscillation modes and dispersion curves.
The methods outlined here are well-suited for such analyses, as the general basis concept and its utilities can be applied to either the $D$ matrices obtained with isolated energy contributions or the kernel matrices themselves.

We consider the main contribution of our work to be presented in Sec. \ref{sec:III}. 
Although the ideas presented here are drawn from well-known linear algebra theory, their application to dynamic matrix theory introduces a new framework for analyzing magnetization dynamics.

This framework goes beyond basic numerical procedures. 
As evidenced by the presented applications, utilizing this unified framework enabled us to address several problems that are not so closely related.
Moreover, as previously explained, the same framework can be used to solve and investigate various problems using the eigensolutions of the kernel matrices. 
Additionally, this framework simplifies the analysis, reformulation, and expansion of numerous analytical approximations.
Many of these, can be conceptualized as a reduction of the dynamic matrix into a $2 \times 2$ matrix from which analytical solutions can be easily obtained (see Sec. \ref{sec:magnonic_crystal} for an example). 
Adopting this perspective facilitates a clearer understanding of the validity and limitations inherent in these approximations, in terms of the properties of the basis functions required to compute the $2 \times 2$ representation of the dynamical matrix.

In conclusion, we have explored discretization procedures applicable to the dynamic-matrix method used to solve the LLG equation in the frequency space. 
The obtained results enhances the comprehension of existing magnetization dynamics techniques and contribute to the formulation and advancement of new analytical and numerical methods for solving the LLG equation in the frequency domain.

\section*{Acknowledgement}
This work was funded by the Carlos Chagas Filho Research Support Foundation of Rio de Janeiro State (FAPERJ), through grants E-26/200.594/2022 and E-26/202.083/2022.

\bibliographystyle{ieeetr}
\bibliography{refs}

\end{document}